\magnification=1100
\vsize=22truecm
\parindent=1.2truecm
\baselineskip24truept
\overfullrule0pt 
\def\cl{\centerline}
\def\nt{\noindent}
\def\bs{\vskip.3in}
\def\ms{\vskip12pt}
\def\ss{\vskip6pt}
\def\ds{\displaystyle}
\font\bb=msbm10

\def\R{\hbox{\bb R}}

\font\med=cmbx10 at 12pt

\cl{\med {EXPLORING THE SIMILARITIES OF THE}}
\ss

\cl{\med {dS/CFT AND AdS/CFT CORRESPONDENCES}}

\ms
\cl{\med{by}}

\ss

\cl{\med {Brett McInnes}}
\baselineskip14truept
\cl{\bf Department of Mathematics}

\cl{\bf National University of Singapore}

\cl{\bf 10 Kent Ridge Crescent}

\cl{\bf Singapore 119260}

\cl{\bf Republic of Singapore}

\cl{matmcinn@nus.edu.sg}

\bs
\baselineskip15truept
\nt {\bf Abstract.}\ \ The $dS/CFT$ correspondence differs from its AdS/CFT counterpart in some ways, yet is strikingly similar to it in many others.  For example, both involve CFTs defined on {\it connected} spaces (despite the fact that the conformal boundary of deSitter space is not connected), and both impose constraints on scalar masses (Strominger's bound for deSitter, and the Breitenlohner-Freedman bound for Anti-deSitter).  We argue that these similarities can be 
explored and exploited using a slight extension of the Euclidean approach to AdS/CFT. The methods are particularly compatible with Hull's embedding of deSitter Space in a timelike T-dual version of M-theory.
\bs
\nt{\bf I.\ \ INTRODUCTION}
\ms
deSitter space [1] does not represent a realistic cosmological model.  However, as the simplest, maximally symmetric space-time with a positive cosmological constant, it undoubtedly has much to teach us about more realistic [2] models.  In that spirit, several attempts have been made [3], [4], [5], [6] to construct a ``$dS/CFT$ correspondence'', after the manner of the $AdS/CFT$ correspondence [7].  The proposal of [6] is particularly concrete, but it has two features that require further investigation.
\ss
The first unusual aspect of the $dS/CFT$ duality is that the $CFT$ in question is defined on a {\it connected} manifold (a sphere with the canonical conformal structure.)  This is unexpected, since the conformal boundary of deSitter space is disconnected: it consists of two spheres, one in the future and one in the past.  In the specific case of deSitter space, this is explained physically [4], [6] by the fact that a null curve originating at past infinity can reach the antipodal point of the sphere at future infinity.  The two spheres are causally related, and so the relevant Green functions respect only one copy of the conformal symmetry group.  Therefore it is natural, in this case, to think of the CFT being defined on one copy of the sphere.  However, a perturbation of deSitter space (caused, for example, by adding a suitable distribution of matter or gravitational radiation) can destroy the symmetries of the space-time, and then it becomes unclear whether a one-to-one correspondence continues to hold (or whether, if it does, correlators with one point at future infinity and one at past infinity can always be interpreted as correlators with both points at past infinity).  One would like to see some evidence, either way, as to whether the (effective) connectedness of the boundary in the $dS/CFT$ correspondence is generic.
\ss
The second interesting peculiarity of the $dS/CFT$ duality is that a detailed computation of the conformal weights corresponding to a scalar field of mass $m$ yields an expression (for four-dimensional deSitter space, $dS_4$)
$$h_\pm = {1\over 2} \left(3 \pm \sqrt{9 - 4m^2L^2} \right), \eqno{(1.1)}$$
where $3/L^2$ is the cosmological constant of $dS_4$.  Thus it appears that masses of scalar particles greater than $3/2L$ would cause the CFT to become non-unitary; this in turn seems to indicate that deSitter quantum gravity admits no stable scalar particles with masses beyond $3/2L$.
\ss
Now as one would expect, the $dS/CFT$ and the $AdS/CFT$ correspondences differ in some ways: for example, as is stressed in [8], $AdS_4$ contains non-normalizable modes (corresponding [9] to boundary {\it sources}), but these do not exist in $dS_4$.  Nevertheless, there are also unexpected and striking similarities, and the two properties of $dS/CFT$ to which we drew attention above provide particularly interesting examples.  It is well known [10] that there are strong physical reasons for {\it requiring} the boundary to be connected in the $AdS/CFT$ case.  Such reasons are not evident in the case of $dS/CFT$, yet the boundary remains (effectively) connected.  Again, the existence of a bound on the masses of scalar particles in $AdS_4$ is well known $-$ while such particles are permitted to be tachyonic, the Breitenlohner-Freedman bound states that they can be stable only if the mass satisfies [11]
$$m^2 \ge -9/4L^2, \eqno{(1.2)}$$
where $-3/L^2$ is the cosmological constant of $AdS_4$.  An $AdS_4$ tachyon of mass $im$, $m > 0$, thus satisfies $m \le 3/2L$, {\it which is precisely Strominger's bound} : a scalar particle in deSitter quantum gravity is like a tachyon in Anti-deSitter space.
\ss
We take these observations as a strong hint that there is some kind of ``duality of dualities'' between certain aspects of $AdS/CFT$ and $dS/CFT$.  Unfortunately, there are serious obstacles to be overcome if we are to investigate this rather natural idea.  The first is technical: the Euclidean methods which have been so useful [9] in the study of $AdS/CFT$ do not seem to give a {\it complete} picture of $dS/CFT$.  The effective use of Euclidean methods in [8] shows that such techniques are undoubtedly still useful here, but the fact remains that the sphere $S^4$, the usual Euclidean representation of $dS_4$, is not a ``bulk''.  That is, being compact, it cannot be the interior of any manifold-with-boundary; so no bulk-boundary $dS/CFT$ duality can even be formulated if $S^4$ is the only legitimate Euclidean version of $dS_4$.  This is not a peculiarity of $dS_4$.  Take any complete Riemannian manifold $M$ with Ricci tensor eigenvalues bounded below by a positive constant $k^2$; then Myers' theorem ([12], page 165) states that $M$ is compact with diameter $\le \pi /k$.  Thus it seems that any cosmology dominated by a positive cosmological constant will have a Euclidean counterpart which cannot be interpreted as a ``bulk''.  (Notice that the restriction on the diameter means that the ``shape'' as well as the topology is constrained.)  All this is of course in sharp contrast to the case of $AdS_4$, for which the Euclidean version is the hyperbolic space $H^4$; the latter is naturally represented as the interior of a closed ball, with boundary $S^3$, and the $AdS/CFT$ duality (in the context, for example, of M-theory on $AdS_4 \times S^7$) can be easily and very usefully formulated in Euclidean language as a duality between theories on $H^4$ and $S^3$. 
\ss
Although Euclidean methods on $S^4$ are undoubtedly useful [8] in studying some aspects of $dS/CFT$ duality, they clearly must be supplemented in some way if we are to understand the similarities between the two dualities.  Our proposal is that Euclidean manifolds of signature ($+ \ + \ + \ +$) should be supplemented by ``negative Euclidean'' manifolds of signature ($- \ - \ - \ -$).  This seemingly trivial extension of the Euclidean technique is of interest because it allows us to find a non-compact complete manifold, of {\it positive} curvature, corresponding to $dS_4$.  By combining the results of the ``positive'' and ``negative'' Euclideanizations, we may be able to construct a complete Euclidean picture of $dS/CFT$ duality.
\ss
A study of the similarities between $AdS/CFT$ and $dS/CFT$ could well be relevant to the notorious problem of ``embedding'' $dS_4$ in $M$-theory [4], [5], [13] [14].  This can be done indirectly, by embedding a deSitter space in an Anti-deSitter space of higher dimension [15] and perhaps unusual topology [16]; but a direct embedding would also be instructive.  The best that has been achieved in this direction is the demonstration by Hull and co-workers (see particularly [3], [17], [18]) that there are deSitter solutions of $M^*$ theory.  This theory is the strong-coupling limit of $IIA^*$ string theory, which is obtained by {\it timelike} {\it T-duality} from $IIB$ theory.  Thus $M^*$ theory has two time dimensions, and has the expected difficulties with ghosts.  If it is true, however, as Hull argues, that $M^*$ theory represents just another (inhospitable) ``corner'' of the moduli space of an underlying fundamental theory, then the fact that $M^*$ solutions do not exactly resemble our world is not a fatal objection.  What it does mean is that the behaviour of particles and fields on deSitter space can only be understood by explicitly constructing a duality between the deSitter solution of $M^*$ theory and a better-behaved solution of $M$ theory.  The problem, of course, is that it is not clear how to do this, particularly without the aid of Euclidean techniques.  We shall argue that ``negative Euclidean'' techniques, which emphasise the similarities between $AdS/CFT$ and $dS/CFT$, may likewise allow us to connect deSitter solutions of $M^*$ theory to Anti-deSitter solutions of $M$ theory.
\bs
\nt{\bf II.\ \ NEGATIVE EUCLIDEANIZATION}
\ms
The ``Euclidean'' approach to path integrals is based on the simple observation that convergence is far easier to achieve if the signature of space-time is taken to be ($+ \ + \ + \ +$) instead of ($ \ - \ + \ + \ +$).  For example, the Lorentzian scalar field Lagrangian is 
$$L(- \ + \ + \ + ) = - \ {1\over 2}\; g^{ij} \nabla_i \phi \ \nabla_j\phi - \; {1\over 2} m^2\phi^2, \eqno{(2.1)}$$
which has no particular sign; but its Euclidean counterpart,
$$L (+ \ + \ + \ +) = \ {1\over 2}\; g^{ij} \;\nabla_i \phi \ \nabla_j\phi + 
\; {1\over 2} m^2\phi^2, \eqno{(2.2)}$$
is evidently non-negative (leaving tachyonic excitations aside).  Of course, had we begun with the ``mostly negative'' signature ($+ \ - \ - \ -$), then the Lagrangian would have been
$$L(+ \ - \ - \ -) = {1\over 2} \; g^{ij} \; \nabla_i \phi \ \nabla_j\phi - 
\; {1\over 2} m^2\phi^2, \eqno{(2.3)}$$
and then ``complexifying time'' would have led to
$$L(- \ - \ - \ -) = -\ {1\over 2} \; g^{ij} \nabla_i \phi \ \nabla_j\phi +  
\; {1\over 2} m^2\phi^2, \eqno{(2.4)}$$
which, with signature ($- \ - \ - \ -$), is positive or zero.
\ss
Obviously one cannot claim that (2.1) is ``more physical'' than (2.3), and the same is surely true of (2.2) and (2.4).  It is a mere matter of convention that Riemannian metrics are taken to be positive rather than negative definite.  (Of course, one has to take the absolute value before taking the square root to compute distances, just as one does for timelike displacements in signature ($- \ + \ + \ +$).)  The reader may feel that the distinction is therefore completely trivial, and that is indeed so in a {\it flat} spacetime. {\it It is otherwise, however, for a curved spacetime}. For example, it is easy to see that the curvature tensor is invariant under a sign reversal of the metric, and hence so is the $(0,2)$ version  of the Ricci tensor.  Thus for example the negative hyperbolic metric
$$g^{-H^4} = -dr \otimes dr - {\sinh}^2 \left( {r\over L}\right) g^{S^3}, \eqno{(2.5)}$$
where $g^{S^3}$ is the (positive) metric on the round 3-sphere of radius $L$, satisfies
$$R^{-H^4}_{ij} = + {3\over L^2} \; g^{-H^4}_{ij}. \eqno{(2.6)}$$
\nt We must therefore think of $-H^4$, the negative hyperbolic space, as a space corresponding to a Lorentzian space-time of {\it positive} cosmological constant.
\ss
From a physical standpoint, it may seem that there is nothing to be gained from this exercise, but this is not so: for two reasons. First, the distinction between the two kinds of Riemannian manifold will certainly be important when ``negative'' manifolds are in some way brought into contact with ``positive'' ones.  For example, Hull's timelike T-duality leads to an M-theory analogue defined on an 11-dimensional manifold of signature (6,5), and in [17] and [18] Freund-Rubin solutions are considered on the (6,5) signature manifold $AdS_7 \times (-H^4)$.  Here $AdS_7$ is the usual anti-deSitter space with signature (6,1) and negative curvature, while $-H^4$, as above, has {\it positive} curvature, so that the two factors, as usual, have opposite signs of curvature.  In this case, the juxtaposition with $AdS_7$ imposes a genuine, physical distinction between $H^4$ and $-H^4$.  It is clear that, particularly in the context of Hull's theories, {\it there can be no justification whatever for preferring positive Euclideanizations to negative ones}.  Thus, if complexification of a Lorentzian metric leads us to a metric of signature $(- \ - \ - \ -$), we should expect this (with matter Lagrangians like (2.4)) to be just as useful as a metric of signature ($+ \ + \ + \ +$).  If this is accepted, however, then we must also allow a given Lorentzian metric to have {\it two} Euclideanizations, one positive, one negative.  The emphasis on one or the other will be determined by the exigencies of each individual problem.  The problem at hand is that Myers' theorem, discussed in the Introduction, forbids a purely positive Euclidean formulation of the $dS/CFT$ correspondence, since it requires the positive Euclidean version of $dS_4$ to be compact.  The advantage of the negative Euclideanization is that it reverses the relationship between curvature and compactness : in negative Riemannian geometry, a complete manifold must be compact if its Ricci eigenvalues are bounded {\it above} by $-k^2$, $k$ real.  A Lorentzian manifold of positive cosmological constant will have a negative Euclideanization of positive scalar curvature, but such a manifold will normally be non-compact and will have a non-trivial conformal boundary. This is the second reason for thinking that negative Euclideanization could be physically interesting. Some examples will clarify the point.
\ss
The round metric on the four-sphere,
$$g^{S^4} = L^2 d\alpha \otimes d\alpha + {\cos}^2(\alpha) g^{S^3}, \eqno{(2.7)}$$
is represented in polar coordinates (we use latitude instead of co-latitude for later convenience) as a local warped product metric on $S^1 \times S^3$.  Similarly it is possible to represent $S^4$ as the local warped product of a pair of two-spheres, $S^2_a$ and $S^2_b$ (see [12], page 272):
$$g^{S^4} = g^{S_{a}^{2}} + {\cos}^2(\alpha) g^{S_{b}^{2}}. \eqno{(2.8)}$$
The co-ordinate singularities here remind us, of course, that $S^4$ is not really $S^2 \times S^2$.  Thus $-S^4$ is a negative Riemannian manifold with metric
$$g^{-S^4} = -L^2(d\alpha \otimes d\alpha + {\sin}^2(\alpha) d\beta \otimes d\beta) - L^2{\cos}^2(\alpha)(d\theta \otimes d\theta + {\sin}^2(\theta) d\phi \otimes d\phi). \eqno{(2.9)}$$
Now complexify by replacing $\alpha$ by $i\alpha - {\pi \over 2}$ to obtain the Lorentzian metric
$$L^2 d\alpha \otimes d\alpha - L^2 {\cosh}^2(\alpha) d\beta \otimes d\beta + L^2 {\sinh}^2(\alpha) (d\theta \otimes d\theta + {\sin}^2(\theta) d\phi \otimes d\phi). \eqno{(2.10)}$$
Defining $t = L\beta$, $r = L\alpha$, we have
$$-{\cosh}^2\left( {r\over L}\right) dt \otimes dt + dr \otimes dr + L^{2}{\sinh}^2 \left({r\over L}\right) (d\theta \otimes d\theta + {\sin}^2 (\theta) d\phi \otimes  d\phi), \eqno{(2.11)}$$
which is {\it precisely the metric of} $AdS_4$ {\it in globally valid coordinates} (see [1], page 131).  Thus we conclude that the negative Euclidean form of $AdS_4$ is $-S^4$, which is compact; this agrees with the negative version of Myers' theorem discussed earlier.
\ss
Next, take $g^{-H^4}$ given by (2.5), and define $\alpha$ by
$$1/{\sinh}(\alpha) = {\sinh}(r/L), \eqno{(2.12)}$$
obtaining
$$g^{-H^4} = {1\over {\sinh}^2(\alpha)} [-L^2 d\alpha \otimes d\alpha - g^{S^3}]. \eqno{(2.13)}$$
From the point of view of any bulk-boundary duality, this is the fundamental form of the metric : for it explicitly reveals the relationship between the bulk geometry and the geometry at infinity, which is evidently that of a negative conformal 3-sphere.  Complexify $\alpha$ to $i\alpha$, and obtain the Lorentzian metric
$${1\over {\sin}^2 (\alpha)} \;\ [-L^2 d\alpha \otimes d\alpha + g^{S^3}]. \eqno{(2.14)}$$
This will seem more familiar if we set $e^{t/L} = {\tan} (\alpha /2)$, for then we have simply
$$-dt \otimes dt + {\cosh}^2 \left({t\over L}\right) g^{S^3}, \eqno{(2.15)}$$
{\it which is precisely the} deSitter {\it metric in globally valid coordinates} (see [1], page 125).  The negative Euclidean version of $dS_4$ is $-H^4$, a complete manifold of positive scalar curvature which is nevertheless not compact : negative Euclideanization circumvents Myers' theorem.
\ss
Thinking of $-H^4$ as a Euclideanization of deSitter space may seem no more than a formal device, but there is much to be learned in this way. (We repeat that we are by no means claiming that $S^4$ is the ``wrong'' Euclideanization of $dS_4$.  Different problems emphasise different Euclideanizations.)  Notice particularly that $dS_4$ and $-H_4$ {\it have the same symmetry group}.  Recall that $dS_4$ can be regarded as a submanifold of $\R^5$, as follows: with the Lorentzian metric
$$ + dT \otimes dT - dt \otimes dt + dx \otimes dx + dy \otimes dy + dz \otimes dz, \eqno{(2.16)}$$
\nt the locus
$$+ T^2 - t^2 + x^2 + y^2 + z^2 = L^2, \eqno{(2.17)}$$
$L$ real, is $dS_4$.  The isometry group is $O(4,1)$.  Similarly, $-H^4$ is a connected component of the locus
$$-T^2 + t^2 + x^2 + y^2 + z^2 = -L^2 \eqno{(2.18)}$$
in $\R^5$ with the metric
$$+ d T \otimes dT - dt \otimes dt - dx \otimes dx - dy \otimes dy - dz \otimes dz, \eqno{(2.19)}$$
where the reversal of signs with respect to the left side of (2.18) gives the negative metric on $-H^4$.  Notice that (2.16) is obtained from (2.19) by complexifying $x$, $y$, and $z$ (a process which does not affect the structure of the isometry group); when this is done, (2.18) becomes (2.17), that is, $-H^4$ becomes $dS_4$ as expected.  Clearly the isometry group of $-H^4$ is again $O(4,1)$.
\ss
Thus we see that it is possible to Euclideanize $dS_4$ in a way {\it that does not change its symmetry group}.  (This is a special property of deSitter space : {\it neither} the positive {\it nor} the negative Euclideanization of $AdS_4$ has the same symmetry group as $AdS_4$).  This fact is again related to the fact that negative Euclideanization allows us to evade the conclusions of Myers' theorem.  As we saw, the positive Euclideanization of a (geodesically complete) Lorentzian Einstein manifold of positive scalar curvature must be compact; but the symmetry group of a compact Riemannian manifold must itself be compact ([19], page 239) and so it can never be a Lorentzian symmetry group.  In other words, $dS_4$ has a non-compact symmetry group because it is Lorentzian, and $-H_4$ can have this same symmetry group because it is not compact.  The real difference between $dS_4$ and $-H_4$ is revealed at the level of their ``gauge groups'', that is, their {\it holonomy} groups.  These can be computed ([12], page 297) by studying the respective subgroups of $O(4,1)$ that fix a point.  Noting that the point $(L,0,0,0,0)$ lies on both (2.17) and (2.18), we see that the holonomy group of $dS_4$ is the identity component of $SO(3,1)$, while that of $-H^4$ is $SO(4)$.  Thus $-H^4$ is better behaved from this point of view: it has a compact ``gauge group''.
\ss
If one wishes to compare the $AdS/CFT$ and $dS/CFT$ correspondences, it is evident that $-H^4$ is the relevant Euclideanization of $dS_4$ : many simple facts about the correspondences are clarified in this way.  Most basically, $-H^4$, unlike $S^4$, can be regarded as the interior of a manifold with boundary, and so a Euclidean version of the correspondence, with $-H^4$ as the bulk and $-S^3$ as its conformal boundary (see (2.13)), can be set up.  The fact that both $AdS/CFT$ and $dS/CFT$ involve conformal field theories on spheres is now understandable.  In the case of $AdS$, only the Euclideanized version actually involves a sphere, while for $dS$ {\it both} the Lorentzian {\it and} the Euclidean versions do so; but we now see that this is related to the fact that $dS_4$ and $-H^4$ have the same symmetry group.  In general, the fact that $H^4$ and $-H^4$ are so similar should lead us to expect that many computations in $dS/CFT$ will yield similar results to their $AdS/CFT$ counterparts.  For example, notice that while the scalar curvature and the $(1,1)$ version of the Ricci tensor reverse sign when the sign of the metric is reversed, their squares are invariant.  Since the formula for the holographic conformal anomaly involves only these squares, one expects the expression for it to be similar in the two correspondences: and so it proves [20].
\ss
With this preparation, we can re-consider the questions, raised in the Introduction, concerning the $dS/CFT$ correspondence.
\bs
\nt{\bf III.\ \ EFFECTIVE CONNECTEDNESS OF THE BOUNDARY IN THE $dS/CFT$ CORRESPONDENCE}
\ms
The basic observation of the previous section was very simple: since the positive Euclideanization of $AdS_4$ (namely $H^4$) is so very similar to the negative Euclideanization of $dS_4$ (namely $-H^4$), the basic versions of the $AdS/CFT$ and $dS/CFT$ correspondences are bound likewise to be very similar. Nevertheless, the fact remains that deSitter cosmology is very different to Anti-deSitter cosmology. In this section, we shall argue that those differences become apparent when one introduces matter into spacetime and allows it to modify the geometry: for the kinds of {\it boundary conditions} one is entitled to impose in the two cases are very different.
\ss
One of the most basic facts hinting at the existence of the $AdS/CFT$ duality is the isomorphism between the isometry group of Anti-deSitter space and the conformal group of (compactified) Minkowski space.  If we try to construct an analogous argument in the case of deSitter space, we find two peculiar facts.  First, this space-time has {\it two} conformal boundaries, each being a conformal 3-sphere with conformal group $O(4,1)$.  Clearly we cannot expect a duality between a boundary with a conformal group consisting of {\it two} copies of $O(4,1)$ and a bulk with a symmetry group isomorphic to {\it one} copy.  The duality must involve one, connected component of the boundary : in other words, the boundary must, from a physical point of view, be {\it effectively} connected.  The second odd feature is that the duality is between a Lorentzian bulk and a Euclidean boundary.  This suggests that a Euclidean formulation may shed some light on this ``effective connectedness''.
\ss
The immediate physical reason for the effective connectedness of the $dS$ boundary derives from the observation [4], [6] that null curves in deSitter space induce an antipodal identification of the $S^3$ at past infinity with the $S^3$ 
at future infinity.  The relevant Green functions therefore cannot be expected to respect two independent copies of the conformal group.
\ss
The interesting feature of these arguments is that they rely on a knowledge of the precise symmetries and causal structure of deSitter space.  Both of these are modified by perturbations.  Witten [4] refers to a result in [21] which states the following.  Let $M$ be any spacetime which is null geodesically complete, which satisfies the null energy and null generic conditions, and which is globally hyperbolic with a compact Cauchy surface.  Then there exist Cauchy surfaces $\Sigma_1$ and $\Sigma_2$ such that for any event in the chronological future of $\Sigma_2$, the chronological past of that event contains $\Sigma_1$.  This statement is false in deSitter space itself, which does not satisfy the null generic condition, for at no (finite) time can a deSitter observer ``see'' an entire Cauchy surface; hence, null curves establish a one-to-one (antipodal) correspondence only between the infinite past and the infinite future.  A generic perturbation (due to gravitational radiation, matter, or quantum fluctuations) of deSitter space {\it will}, however, satisfy the conditions of the theorem.  This means that the re-focusing of the null geodesics emanating from a point at past infinity (independent of their original direction) will, in a generically perturbed deSitter space, take place {\it before} future infinity is reached. Can some effect of this kind break the effective identification of the two disconnected components of deSitter infinity?  The question could be very important, since breaking the identification would produce a non-singular past-future correlator, and Witten argues [4] that this could render the deSitter Hilbert space finite-dimensional, as suggested by holographic arguments [22]. However, it is clear that we do not expect {\it every} perturbation (or {\it all} kinds of matter) to break the effective identification. Therefore, this argument is merely a hint that perturbations or matter of {\it some} kind do break it $-$ the question remains: can we identify these perturbations more precisely?

\ms
For Anti-deSitter holography, questions about the connectedness of the boundary are usually considered in the Euclidean domain [10], [23].  It is reasonable to hope that a Euclidean approach could be instructive in the deSitter case also, but first we must be clear as to the meaning of the results of [10] and [23].
\ms
The main result of [10] and [23], which we call the WYCG theorem, states the following.  Let $W^{n+1}$ be a complete, connected Riemannian asymptotically hyperbolic manifold with conformal boundary $N^n$ and asymptotic sectional curvature $-1/L^2$.  Suppose that the eigenvalue functions of the Ricci curvature satisfy $\lambda_\beta \ge -n/L^2$ and that they approach $-n/L^2$ asymptotically in such a way that, with a canonical choice of defining function $F$ (that is, $F=0$ ``defines'' the boundary),
$$F^{-4}(\lambda_\beta + {n\over L^2}) \to 0 \eqno{(3.1)}$$
uniformly as the boundary is approached.  Suppose finally that the conformal structure induced on any connected component of $N^n$ is represented by a metric of constant, non-negative scalar curvature.  Then $N^n$ must be connected.
\ms
The WYCG theorem is strictly ``Euclidean''.  Its {\it Lorentzian} interpretation is {\it not} that boundaries of asymptotically Anti-deSitter space-times must be connected.  Euclidean techniques are a device for understanding the nature and structure of partition functions.  The WYCG result means that, as far as the partition function is concerned $-$ that is, for all physical purposes $-$ we may as well treat the boundary {\it as if it were} connected (provided that the conditions of the theorem are satisfied).  This interpretation, sketched in [24], was confirmed by [25], where it was shown that, under physically reasonable assumptions, the various connected components of the conformal boundary of an asymptotically Anti-deSitter spacetime cannot communicate with each other $-$ they are {\it causally} disconnected.  From the AdS/CFT point of view, this implies that bulk physics in an asymptotically Anti-deSitter spacetime must be completely encoded holographically on {\it each} connected component of the boundary, separately : for the information on a given connected component cannot be supplemented by information stored on other connected components.  But then it follows that these other connected components are superfluous.  That is, the WYCG theorem means that the Lorentzian boundary is {\it effectively} connected.
\ms
The conditions assumed in the WYCG theorem all have a physical significance.  Condition (3.1) requires matter fields to decay rapidly towards spatial infinity [26], while $\lambda_\beta \ge - n/L^2$ is essentially a positivity condition on matter potentials; finally, the demand that the scalar curvature at infinity should be non-negative is a stability condition for the CFT there [27].  Under these conditions, then, the Euclidean path integral will only involve connected boundaries, a procedure with a physical interpretation given above.
\ms
Now as we have seen, [4] and [6] indicate that the conformal boundary of deSitter space is also {\it effectively} connected.  This is reflected in the structure of the negative Euclideanization of $dS_4$, since $-H^4$ has a single copy of $-S^3$ as its conformal boundary.  The matching of the symmetry group of $dS_4$ with that of {\it one} conformal sphere is now explicable, since $dS_4$ and $-H^4$ have precisely the same symmetry group.  These simple observations suggest to us that a (generalized) Euclidean path integral can be relied upon to detect effective connectedness in the deSitter case as well as in the Anti-deSitter case.  Let us set up a (necessarily negative) Euclidean bulk-boundary correspondence for asymptotically deSitter spaces, and see what we can learn about the connectedness of the boundary in this case.
\ms
$-H^4$ is an example of a class of negative Riemannian manifolds which can be described as follows.  Let $W^{n+1}$ be a non-compact $(n+1)$-dimensional manifold which can be regarded as the interior of a compact, connected manifold-with-boundary ${\overline W}^{n+1}$, and let $N^n$ be the boundary (which need not be connected).  Let $g^{-W}$ be a smooth negative metric on $W^{n+1}$ such that there exists a smooth function $F$ on ${\overline W}^{n+1}$ with the following properties:
\ss
\item{(i)}		$F(x) = 0$ if and only if $x \in N^n$;

\item{(ii)}	$dF(x) \ne 0$ for all $x \in N^n$;

\item{(iii)}	$F^2g^{-W}$ extends continuously to a negative metric on 
			${\overline W}^{n+1}$;

\item{(iv)}	If $|dF|_F$ is the norm of $dF$ with respect to the extended 			negative metric, then $|dF|_F$, evaluated on $N^n$, must not 			depend on position there.

\ss
\nt Then $-W^{n+1}$ (that is, $W^{n+1}$ endowed with $g^{-W}$) may be called an {\it asymptotically negative hyperbolic space}; for all sectional curvatures along geodesics ``tending to infinity'' approach a common {\it positive} constant, which we may take to be $1/L^2$.  The negative metric $g^{-W}$ induces a negative conformal structure on $-N^n$, represented by a negative metric of constant scalar curvature, which may be of any sign.  For the standard negative metric (2.13) on $-H^4$, the sectional curvatures are all equal to $+1/L^2$ everywhere, but the induced negative conformal structure at infinity is represented by the canonical metric on $-S^3$,  with scalar curvature $-6/L^2$.
\ss
The WYCG theorem translates straightforwardly into ``negative'' language as follows.  Let $-W^{n+1}$ be a connected complete asymptotically negative hyperbolic manifold with asymptotic sectional curvature $+1/L^2$.  The eigenvalue functions of the $(1,1)$ version of the Ricci tensor are then asymptotic to $n/L^2$.  Suppose that, with a canonical choice of defining function $F$,
$$F^{-4}\left(\lambda_\beta - {n\over L^2}\right) \to 0 \eqno{(3.2)}$$
uniformly as the boundary is approached.  Suppose too that
$$\lambda_\beta \le {n \over L^2} \eqno{(3.3)}$$
for all $\beta$ and that the negative conformal structure induced on any connected component of the conformal boundary $-N^n$ is represented by a (negative) metric of non-positive scalar curvature.  Then $-N^n$ is connected.
\ss
Obviously $-H^4$ and its conformal boundary $-S^3$ satisfy the conditions and the conclusion of this theorem.  More generally, an asymptotically deSitter spacetime, as considered in [4], will have a negative Euclideanization which is asymptotically negative hyperbolic.  The stability of the boundary $CFT$ depends on the spectrum of a conformally invariant operator of the form [10]
$$\nabla^2 + {n-2 \over 4(n-1)}R \eqno{(3.4)}$$
where $\nabla^2$ is the Laplacian and $R$ is the scalar curvature.  Since the Laplacian reverses sign under sign reversal of the metric, the stability condition is that the negative conformal structure on the boundary should be represented by a metric of non-positive scalar curvature; so this assumption in the ``negative WYCG'' theorem has a physical rationale.
\ss
If we now consider asymptotically deSitter metrics such that the bulk remains precisely Einstein, then each $\lambda_\beta$ is precisely $\ds{n\over L^2}$, and so conditions (3.2) and (3.3) are satisfied trivially.  Therefore, if the bulk is connected and complete, {\it the Euclidean boundary remains connected}, no matter how different the metric may be from the deSitter metric (away from the boundary).
\ss
The physical interpretation of this result is as follows.  We can introduce perturbations of deSitter space which destroy its symmetries and yet preserve the Einstein condition : for example, gravitational waves or certain quantum fluctuations can have this effect.  Such perturbations maintain a connected Euclidean boundary, so we expect the Lorentzian boundary to continue to be effectively connected.  In short, the assumption in [6] that the $dS/CFT$ correspondence should only involve one connected component of the conformal boundary appears to be generically valid in this sense; that is, the fact that deSitter space is so symmetrical (in fact, it is maximally symmetric) is not crucial.
\ss
Before accepting this conclusion, however, we should carefully examine our assumptions.  In particular, we have only considered perturbations that maintain the Einstein condition; but it would be more natural to consider also a matter field which affects the deSitter geometry through the (assumed approximately valid) Einstein equation.  In that case we must consider whether the conditions (3.2) and (3.3) can still be expected to hold.  For example, we can consider a scalar field as in [13] and [14], with a Lagrangian generalizing (2.4) by the inclusion of a {\it positive} (typically exponential) potential, 
$$L^V(- \ - \ - \ - ) = - {1\over 2} \; g^{ij} \; \nabla_i \phi \; \nabla_j \phi + V(\phi), \eqno{(3.5)}$$
with a negative-Euclidean stress-energy-momentum tensor
$$T_{ij} = \nabla_i\phi \; \nabla_j\phi - {1\over 2} \; g_{ij} (\nabla_k\phi \nabla^k\phi) + g_{ij} \; V(\phi). \eqno{(3.6)}$$
Using Einstein's equation (in the four-dimensional case) in the form $R_{ij} - \ds{1\over 2} \; g_{ij} R + \ds{3\over L^2} \; g_{ij} = T_{ij}$, we have
$$R_{ij} - {3 \over L^2} g_{ij} = \nabla_i\phi \; \nabla_j\phi - g_{ij} \; V(\phi). \eqno{(3.7)}$$
If $t^i$ is   a unit eigenvector of the Ricci tensor $(t^it_i = -1)$ with eigenvalue $\lambda$, we find
$$\lambda - {3 \over L^2} = - (t^i \nabla_i\phi)^2 - V(\phi), \eqno{(3.8)}$$
so (3.3) is indeed satisfied provided that, as is normally the case, the potential is non-negative.
\ss
Condition (3.2) is, however, more questionable.  It may be reasonable, particularly in five or more dimensions [16], to ask for such a rapid rate of decay of energy density and pressure towards {\it spatial} infinity; but here we are concerned with decay of densities and pressures due to cosmological expansion (in the remote past and future).  With energy density $\rho$, pressure $p$, and equation of state $p = w\rho$, the vanishing of the divergence of the stress-energy-momentum tensor yields (in four dimensions)
$$\rho = K a^{-3(1+w)}, \eqno{(3.9)}$$
where $K$ is a constant and $a(t)$ is the scale factor.  For ordinary matter, $w = 0$; for relativistic matter, $w = \ds{1\over 3}$; and for quintessence, $w$ is asymptotically negative.  Now the eigenvalues of the Ricci tensor (minus the cosmological constant) are, by the Einstein equation, related to the energy density and the pressures, and $a(t)$ plays the role (when the metric is expressed conformally) of $1/F$.  Even in the case of relativistic matter, the density and pressure are proportional to $a^{-4}$, so we can only justify a requirement that $F^{-4}\left(\lambda - {3\over L^2}\right)$ should tend to a (non-zero) constant. In fact, recent observations [28] hint at a value of $w$ slightly larger than $-1$. If, currently, $w = -0.8$, then $\rho$ is currently decaying as slowly as $a^{-0.6}$. The point is this: matter having a density with a very slow rate of decay, while somewhat exotic, is not currently regarded as unphysical. If we superimpose such matter on an ordinary (positive) cosmological constant,
we shall obtain a spacetime which approaches deSitter space surely but very slowly. In short, {\it classically} (though possibly not quantum mechanically $-$ see below), very slow rates of approach to the asymptotic geometry are more acceptable in a deSitter-like spacetime than in an Anti-deSitter-like spacetime. {\it This is where the real physical differences between $dS$ and $AdS$ are reflected in the Euclidean formulation}. For it is no longer clear that we are entitled to impose (3.2). 

\ss
As is pointed out in [10] and [23], the sign condition on the scalar curvature at infinity cannot be relaxed (see below).  Now in the approach of [23], this condition is only needed to ensure that the mean curvature of a certain sequence of compact hypersurfaces should decay in a particular way as the hypersurfaces ``tend to'' infinity; and in fact it is just the failure of the metric to satisfy this decay condition that allows the conformal boundary to be disconnected in examples like (3.11), below.  In other words, it is the mean curvature condition that is essential, not the sign of the scalar curvature at infinity.  But in fact (3.2) is also used to establish this mean curvature condition.  Now it is probable that (3.2) can be weakened, perhaps (we conjecture $-$ no counterexamples are known to the author) even to the extent that the exponent $-4$ can be replaced by $-3$.  But it is clear that completely dropping (3.2) will cause the WYCG result to fail.  Some examples may be helpful at this point.
\ss
A quick, if potentially confusing, proof that $-H^4$ is the negative Euclideanization of $dS_4$ runs as follows.  One can foliate $-H^4$ with copies of $-H^3$ and obtain the metric in the form 
$$g^{-H^4} = - dt \otimes dt + {\cosh}^2\left({t\over L}\right) g^{-H^3}. \eqno{(3.10)}$$
Now recall that $g^{-H^3}$ is a metric with a {\it positive} cosmological constant $+ 2/L^2$.  Of course, $S^3$ is also a manifold with a positive cosmological constant $+ 2/L^2$.  Replacing $g^{-H^3}$ in (3.10) by $g^{S^3}$, we obtain the (Lorentzian) deSitter metric (2.15).  The structure of (3.10) suggests at first glance that the conformal boundary of $-H^4$ is two copies of $-H^3$, but this is because (3.10) does not represent a conformal {\it compactification} of $-H^4$, since the boundary is not compact.  However, we can still use (3.10) if we replace $H^3$ by $H^3/\Gamma$, where $\Gamma$ acts freely and properly discontinuously and in such a way that $H^3/\Gamma$ is compact (see [29]).  Then the metric
$$-dt \otimes dt + {\cosh}^2 \left( {t\over L}\right) g^{-H^3/\Gamma} \eqno{(3.11)}$$
is {\it locally} identical to the metric of  $-H^4$, so it satisfies (3.2) trivially.  (It no longer corresponds to $-H^4$ globally, however.)  Clearly conformal infinity is disconnected for this manifold, and it is shown explicitly in [23] that this is due to the failure of the mean curvature decay condition; and this in turn is due to the fact that $-H^3/\Gamma$ has positive scalar curvature.
\ss
The alternative way for the mean curvature decay condition to fail occurs if (3.2) is not satisfied.  An easy (but very special) way to construct examples is to modify (3.11).  Consider the manifold $\R \times T^3$, where $T^3$ is the 3-torus, and impose the negative metric
$$-dt \otimes dt + {\cosh}^2\left( {t\over L}\right) g^{-T^3}, \eqno{(3.12)}$$
where $g^{-T^3}$ is a flat negative metric on the torus.  This is not an Einstein metric: the $(1,1)$ components of the Ricci tensor are
$$\eqalignno{
R^t_t &=  + 3/L^2  &(3.13) \cr
R^\alpha_\beta &= \left( {3\over L^2}\right) \delta^\alpha_\beta - 2sech^2 \left({t\over L}\right) \delta^\alpha_\beta , \; \; \alpha, \beta = 1,2,3.  &(3.14)\cr}$$
Clearly, the scalar curvature at infinity is non-positive, and (with $n  =3$) condition (3.3) is satisfied, but (3.2) is {\it not}.  That is, the rate at which the Ricci eigenvalues tend to $3/L^2$ is too slow for the (negative) WYCG theorem to be used, and this is how the manifold with metric (3.12) can have a disconnected conformal boundary.
\ss
To summarise: the WYCG theory requires conditions on the sign of the scalar curvature at infinity and on the rate at which the Ricci eigenvalues approach their asymptotic value.  Example (3.11) shows that the first condition cannot be weakened, and (3.12) shows that {\it some} condition like (3.2) is needed: the problem is to understand the physical significance of slower rates of approach to deSitter density and pressure.
\ss
Now of course (3.11) and (3.12) do not correspond to realistic cosmological models, but this is not relevant $-$ the same can be said of deSitter space itself.  Rather, we must ask whether string/M theory makes sense on such backgrounds.  We saw earlier that this is {\it not} the case for (3.11), since the conformal field theory on the boundary is not stable for such a background (and the bulk theory is unstable to the emission of ``large branes'' [27]).  But this objection does not apply to (3.12) or to its Lorentzian version
$$-dt \otimes dt + {\cosh}^2 \left( {t\over L}\right) g^{T^3}. \eqno{(3.15)}$$
For large positive and negative $t$, this resembles deSitter space (locally) very closely, since a large torus is locally indistinguishable from a large sphere; and while the rate at which it comes to resemble deSitter space may be slow by the standards of the WYCG theory, it is not immediately apparent why we should take exception to it on physical grounds.  (Indeed, in view of the observational evidence favouring flat spatial sections, (3.15) might even be preferred to deSitter space for moderately large positive $t$.)  In short, (3.15) appears to be an example of a background for string theory such that the negative Euclideanization has a disconnected conformal boundary.  This should correspond, in the Lorentzian theory, to a breaking of the effective identification of the two components of conformal infinity.  How does the matter content of the space-time represented by (3.15) perform this feat?
\ss
In fact, while (3.15) resembles deSitter space at large positive and negative $t$, its matter content behaves in a bizarre way near $t = 0$.  Using the Einstein equation, we find (see [1], page 136) that the energy density and pressure corresponding to (3.15) are
$$\eqalignno{
\rho &= {3\over L^2} - {3\over L^2} {\rm sech}^2 \left( {t\over L}\right),  &(3.16) \cr
p &= -{3\over L^2} + {1\over L^2} {\rm sech}^2 \left( {t\over L}\right).  &(3.17) \cr}$$
Recall that $\rho = \ds{3\over L^2}$ and $p = -\ds{3\over L^2}$ for deSitter space.  Now clearly (3.16) and (3.17) violate both the strong and the weak energy conditions ([1], pages 89 and 95), but this observation has very little force here: deSitter space itself violates the strong energy condition, and, in general, string theory routinely violates energy conditions [30].  However, (3.16) behaves in a particularly curious way: the energy density is {\it increasing} after $t = 0$, despite the fact that the universe is expanding.  This is characteristic of spacetimes which violate the {\it dominant energy condition} ([1], page 91),
$$\rho \ge |p|, \eqno{(3.18)}$$
since the vanishing of the divergence of the stress-energy-momentum tensor, 
$${\mathop{\rho}\limits^\cdot} + 3(\rho + p){\mathop{a}\limits^\cdot}/a = 0, \eqno{(3.19)}$$
where $a$ is the FRW scale factor and a dot denotes the time derivative, implies that $\rho$ must increase in an expanding, negative-pressure cosmology which violates (3.18). In principle, this should occasion no more surprise than the fact that the density is constant in deSitter space, which is on the brink of violating (3.18).
\ss
More generally, motivated by the structure of condition (3.2), we can consider an asymptotically deSitter  metric of the form
$$-dt \otimes dt +  a(t)^2 g^{T^3}  \eqno{(3.20)}$$
with density and pressure  given asymptotically by
$$\eqalignno{
\rho &= {3\over L^2} - \alpha a^{-\gamma} 			&(3.21)\cr
p  &= - {3\over L^2} + \beta a^{-\gamma},			&(3.22)\cr}$$
where $\alpha$, $\beta$ and $\gamma$ are constants.  Here $\gamma > 0$; since we are interested in violations of the dominant energy condition, we want $\rho$ to increase as $a(t)$ increases, so we take $\alpha > 0$; and since a sensible equation of state will require $|p|$ to increase as $\rho$ increases, $\beta > 0$ also.  Substituting (3.21) and (3.22) into (3.19), we find
$$\gamma = 3(1 - (\beta/\alpha)).\eqno{(3.23)}$$ 
Combining (3.21) and (3.22) with (3.18), we find that the condition for the dominant energy condition to be violated is that $\alpha > \beta$, which is in fact just $\gamma > 0$.  Evidently $\gamma < 3$, which evades (3.2) and corresponds to the fact that the negative Euclideanization of (3.20) has a disconnected conformal boundary.
\ss
We conclude that asymptotically negative hyperbolic manifolds with disconnected conformal boundaries correspond to asymptotically deSitter space times containing, far from the boundary (that is, concentrated near the time of maximum compression) matter which violates the dominant energy condition (DEC). But how might such a violation disrupt the effective identification of future infinity with past? The DEC is mainly used in black hole theory, as for example in the positive mass theorem [31] and in discussions of cosmic censorship ([32], page 305). However, in fact the DEC plays a more basic role. As is well known, the interpretation of the vanishing of the divergence of the stress-energy-momentum tensor $T$ as a (global) conservation law is problematic in a general curved spacetime; but one can show ([1], page 94) that if $div T = 0$ and $T$ satisfies the dominant energy condition, then the vanishing of $T$ on a closed, achronal set implies its vanishing in the domain of dependence of that set. Conversely, in a spacetime with matter violating the DEC, {\it one must not expect conditions at a given event in the far future to be completely determined by events in its chronological past.} In short, matter which violates the DEC should indeed be expected to break the effective identification of future infinity with past. 
\ss
Does this imply that, in the presence of DEC-violating matter, the $dS/CFT$ correspondence necessarily involves a duality between deSitter quantum gravity and two, independent conformal field theories? Perhaps, but note that it is argued in [6] and [33] (see also [34]) that the $dS/CFT$ correspondence itself imposes conditions on the rate at which an asymptotically deSitter metric must approach the deSitter metric towards infinity. Thus far, these conditions have only been formulated in three spacetime dimensions (and not in a fully invariant way), but it is plausible that in four dimensions their effect is precisely to forbid violations of the DEC. (Some indirect evidence pointing in this direction may be found in [35].) If that is so, then it means that the internal consistency of the $dS/CFT$ correspondence {\it requires} the CFT to be formulated on a connected space. It would also mean that slow rates of approach to an asymptotic deSitter spacetime $-$ which, as we saw earlier, are permitted {\it classically} $-$ are forbidden by deSitter quantum gravity.   
\ss
Let us summarize.  The WYCG theory shows that the conformal boundary of the negative Euclideanization of an asymptotically deSitter space-time must remain connected, if the $CFT$ at infinity is stable, as long as the space-time remains precisely Einstein.  Therefore, the effective connectedness of the conformal boundary of deSitter space, noted in [6], does not depend in an essential way on the symmetries of the space-time.  However, the introduction of matter, with density and pressure approaching the deSitter values sufficiently slowly, can break the effective connectedness of the boundary.  Such matter tends to violate the dominant energy condition, and it seems reasonable to suppose $-$ in view of the physical interpretation of this condition $-$ that it affects the evolution of states in way that does indeed break the precise correspondence between the infinite past and the infinite future.  Notice here that ``matter'' in string theory includes such items as classical corrections to the action corresponding to massive string states: these give rise to terms quadratic in the curvature tensor, which contribute {\it effective} stress-energy-momentum tensors that cannot be expected always to satisfy the usual conditions $-$ including the dominant energy condition. On the other hand, the finiteness of the Brown-York stress-energy-momentum tensor, discussed in [6],[33],[34] as a requirement for the $dS/CFT$ correspondence to work, might well forbid violations of the DEC, so that the boundary remains effectively connected in all cases.
\ss
We have argued that  negative Euclideanization sheds light on one aspect of the $dS/CFT$ correspondence.  Now we shall argue that it performs the same service for Strominger's mass bound.
\bs
\nt{\bf IV.\ \ STROMINGER'S MASS BOUND AS A BREITENLOHNER - FREEDMAN BOUND}
\ms
The most serious impediment to a real understanding of the $dS/CFT$ correspondence is of course the problem of obtaining deSitter space as part of a solution of some string or $M$ theory.  The most natural approach to achieving this goal appears to be through Hull's ``timelike duality'' [3].  In particular, $M^*$ theory, an $M$-like theory defined on an 11-dimensional manifold with two timelike directions, has a membrane - like solution of the form
$$H^{-2/3}_2 [dx^2_1 + dx^2_2 + dx^2_3] + H^{1/3}_2 [-dt^2 - (dt')^2 + dy^2_4 + ... + dy^2_9], \eqno{(4.1)}$$
where the coordinates can be restricted so that $-t^2 - (t')^2 + y^2_4 + ... + y^2_9 \le 0$ and where $H_2$ is ``harmonic'' in these transverse coordinates.  The asymptotic geometry is that of $dS_4 \times AdS_7$.  Compare this with the familiar $M2$-brane solution of the normal supergravity equations,
$$H^{2/3}[-dt^2 + dx^2_1 + dx^2_2] + H^{1/3} [dy^2_3 + ... + dy^2_{10}], \eqno{(4.2)}$$
which of course has asymptotic geometry $AdS_4 \times S^7$.  Now observe that this $M$-theoretic solution, when compactified on a Lorentzian torus $T^{(2,1)}$ of signature $(+ \ + \ -)$, is equivalent to (4.1) compactified on a torus $-T^{(2,1)}$, with {\it negative} metric of signature $(- \ - \ +)$.  This relationship between $M^*$ theory and $M$ theory holds in general, and it leads Hull [36] to argue that $M^*$ is fundamentally {\it equivalent} to $M$, in the usual sense: $M^*$ is just another ``corner'' of the moduli space of the underlying universal theory.  In this sense, we can claim to have exhibited $dS_4$ as part of a ``solution'' of that theory.
\ss
Unfortunately, even if this is so, it is not clear how we can use this representation of $dS_4$.  One is naturally disinclined to trust any conclusions about stability or unitarity obtained by thinking of $dS_4$ as an ``$AdS_7$ compactification'' of any theory.  Perhaps, however, we could hope to gain an understanding of {\it some aspects} of quantum gravity on $dS_4 \times AdS_7$ by establishing some kind of link with $AdS_4 \times S^7$.  The fact that (4.1) and (4.2) are related by compactifying on tori with metrics of {\it opposite signs} suggests a role for a negative Euclidean approach.  The hope is that {\it by combining} $AdS/CFT$ {\it with} $dS/CFT$, {\it we can use} $AdS_4 \times S^7$ {\it to learn something about} $dS_4 \times AdS_7$.
\ss
As we know from Section II, $dS_4$ and $H^4$ have precisely the same symmetry group, $O(4,1)$, and the past conformal boundary of $dS_4$ has the same topology and conformal structure as the conformal boundary of $H^4$.  It therefore makes sense to {\it identify} the past conformal boundary of $dS_4$ with the conformal boundary of $H^4$.  This identification unites two manifolds-with-boundaries into a single space with a distinguished hypersurface, with the conformal structure of $S^3$, which is infinitely far from all other points.  Any isometry of either $dS_4$ or $H^4$ extends to a symmetry of the whole space, which we denote by $dS_4 \cup_{S^3} H^4$; so we can say that the symmetry group of $dS_4 \cup_{S^3} H^4$ is also $O(4,1)$.
\ss
Now to construct a Euclidean counterpart of this space, we have to choose a Euclidean version of $dS_4$ that has a non-empty conformal boundary: that is, we must choose the negative Euclideanization, $-H^4$.  But since the conformal boundary of $-H^4$ is $-S^3$, we have to use a slight generalization of the usual definition of a conformal structure, one where the metric is defined modulo scalar multiples which are {\it either} strictly positive or strictly negative.  We shall denote $S^3$ with this structure by $\pm S^3$.  The Euclidean version of $dS_4 \cup_{S^3} H^4$ is thus $-H^4$ identified with $H^4$ along $\pm S^3$.
\ss
We can easily extend this construction.  We saw in Section II that the negative Euclidean version of $AdS_7$ is just $-S^7$, so the negative Euclideanization of $dS_4 \times AdS_7$ is $(-H^4) \times (-S^7) = -(H^4 \times S^7)$, while the positive Euclideanization of $AdS_4 \times S^7$ is of course $H^4 \times S^7$.  Thus, as suggested by Hull's analysis of (4.1) and (4.2), the Euclidean versions of $dS_4 \times AdS_7$ and $AdS_4 \times S^7$ are related by reversing the sign of the metric.  As before, we can sew together $H^4 \times S^7$ and $-(H^4 \times S^7)$ along $\pm S^3$, the conformal boundaries of $H^4$ and $-H^4$.
\ss
Now as can be seen from the Lagrangian given by (2.4), the equation governing a scalar particle of real mass $m$ on $-H^4$ (Euclidean $dS_4$) is
$$\nabla^2 \phi + m^2\phi = 0. \eqno{(4.3)}$$
However, the theory on $\pm S^3$ does not ``know'' whether this is an ``ordinary'' particle on the $-H^4$ side, or a tachyon of mass $im$ on the $H^4$ (Euclidean $AdS_4$) side.  It therefore imposes the usual condition for the stability of a tachyon on $AdS_4$, the Breitenlohner-Freedman bound [11] (see [9] for a Euclidean derivation)
$$(im)^2 \ge -9/4L^2;	\eqno{(4.4)}$$
that is, a scalar field of mass $m$ on $dS_4$ can be stable only if
$$m \le 3/2L, \eqno{(4.5)}$$
where $3/L^2$ is the cosmological constant.  This is precisely Strominger's bound [6] (see also [37] and [38]), derived in the context of the $dS/CFT$ correspondence (provided that we make the assumption that the $CFT$ is unitary).
\ss
Naturally, we do not claim that the above argument constitutes a ``proof'' of Strominger's bound.  Instead it should be regarded as a hint that the $AdS/CFT$ and $dS/CFT$ correspondences are linked in a deep way, and that negative Euclideanization has a role to play in exploring that link.
\bs
\nt{\bf IV.\ \ CONCLUSION}
\ms
Euclidean methods are sometimes viewed with some suspicion : they work, often extremely well (for example, [8], [9], [10], [23]), but it seems fair to say that it is not altogether clear why this is so.  In view of this, it is wise to keep an open mind as to the details of how the Euclidean philosophy should be applied.
\ss
The essence of Euclideanization is the idea that, through complexification, an indefinite space-time signature, $(+ \ + \ + \ -)$, can be replaced by a definite signature : $(+ \ + \ + \ +)$ or $(- \ - \ - \ -)$.  At first sight, it seems that the choice is immaterial, but this is not so.  The fundamental reason is that, since the Ricci tensor is invariant under a sign reversal of the metric, the Ricci eigenvalues (the eigenvalues of the $(1,1)$ version of the tensor) are {\it not} $-$ they reverse sign.  The negative Euclidean version of a space-time (like deSitter space) with a positive cosmological constant can therefore have positive Ricci eigenvalues and yet be non-compact and complete.  This seems to be the only way to circumvent Myers' theorem; and circumvented that theorem must be, since a compact manifold cannot be a ``bulk'' in the sense of the $AdS/CFT$ and $dS/CFT$ dualities.
\ss
Euclidean methods give a deep insight into the effective connectedness of the boundary in the $AdS/CFT$ correspondence [10], [23].  We have argued that (negative) Euclidean methods can be used to investigate the effective connectedness of the boundary in the $dS/CFT$ case [4], [6].  Although the relevant Euclidean versions of $AdS_4$ and $dS_4$ are closely similar, the physical differences become apparent when matter is introduced and its effects on the background are studied.  In particular, the kinds of boundary conditions we are entitled to impose are quite different, since, for example, a conserved positive energy is not to be looked for in deSitter space.  Our investigation suggests that an asymptotically deSitter spacetime can have a negative Euclideanization with a disconnected boundary, but only if the dominant energy condition is violated.  We have argued that such violations might well be expected to disrupt the effective identification of future with past infinity noted in [4] and [6]. However, it seems quite possible that the $dS/CFT$ correspondence itself prevents this from happening. Clearly this point invites further investigation. 
\ss
The negative Euclideanizations of asymptotically deSitter spacetimes are obviously relevant to the particular concerns of this work, but the {\it positive} Euclideanizations (such as $S^4$ in the case of $dS^4$ itself) are important for other applications [8].  It remains an interesting problem to combine the two in some non-trivial way, perhaps through some kind of generalized ``sum over Euclideanizations''. Such an approach may be needed to obtain a full Euclidean formulation, for example, of the deSitter brane-worlds embedded in higher-dimensional deSitter spaces investigated in [39]. (It could be interesting to take the positive Euclideanization of the brane, combined with the negative Euclideanization of the bulk, exploiting the fact that $H^{5}$ is naturally foliated by $S^{4}$.)

\bs
\nt{\bf REFERENCES}
\ms
\item{[1]} S. W. Hawking, G. F. R. Ellis, The Large Scale Structure of Space-Time, 		Cambridge University Press, 1973.
\item{[2]} A. G. Riess et al, The Farthest Known Supernova: Support for an 		Accelerating Universe and a Glimpse of the Epoch of
	     Deceleration, astro-ph/0104455
\item{[3]} C.M.Hull, Timelike T-Duality, deSitter Space, Large N Gauge Theories 		and Topological Field Theory, JHEP 9807 (1998) 021,
	     hep-th/9806146.
\item{[4]} E. Witten, Quantum Gravity In deSitter Space, hep-th/0106109.
\item{[5]} V. Balasubramanian, P. Horava, D. Minic, Deconstructing deSitter,  		JHEP 0105 (2001) 043, hep-th/0103171.
\item{[6]} A. Strominger, The $dS/CFT$ Correspondence,  hep-th/0106113
\item{[7]} O. Aharony, S.S. Gubser, J. Maldacena, H. Ooguri, Y. Oz, Large N Field 		Theories, String Theory and Gravity,
	     Phys.Rept. 323 (2000) 183-386, hep-th/9905111.
\item{[8]} A. J. Tolley, N. Turok,  Quantization of the massless minimally coupled 		scalar field and the $dS/CFT$ correspondence,
	     hep-th/0108119.
\item{[9]} E. Witten, Anti-deSitter Space and Holography, Adv. Theor. Math. Phys. 		2 (1998) 253,  
		  hep-th/9802150.
\item{[10]}E. Witten, S. T. Yau, Connectedness of the Boundary in the $AdS/CFT$ 		Correspondence, Adv.Theor. Math. Phys. 3 (1999) 1635, hep-th/9910245.
\item{[11]} P. Breitenlohner, D. Z. Freedman, Stability in Gauged Extended 		Supergravity, Ann. Phys. 144 (1982) 249.
\item{[12]}A. L. Besse, Einstein Manifolds, Springer-Verlag, 1987.
\item{[13]} S. Hellerman, N. Kaloper, L. Susskind, String Theory and Quintessence, 		JHEP 0106 (2001) 003, hep-th/0104180.
\item{[14]}W. Fischler, A. Kashani-Poor, R. McNees, S. Paban, The Acceleration of 		the Universe, a Challenge for String Theory,  JHEP 0107 (2001) 003, 		hep-th/0104181.
\item{[15]} S. Nojiri, O. Obregon, S. D. Odintsov, (Non)-singular brane-world 		cosmology induced by quantum effects in d5 dilatonic gravity, Phys.Rev. 		D62 (2000) 104003, hep-th/0005127
\item{[16]}  B. McInnes, A Positive Cosmological Constant in String Theory Through 		$AdS/CFT$ Wormholes,
         Nucl.Phys. B609 (2001) 325-343, hep-th/0105151.
\item{[17]}  C.M. Hull, R.R.Khuri, Branes, Times and Dualities, Nucl.Phys. B536 		(1998) 219, hep-th/9808069.
\item{[18]}  C.M. Hull, R.R.Khuri, Worldvolume Theories, Holography, Duality and 			Time, Nucl.Phys. B575 (2000)
          231, hep-th/9911082.
\item{[19]}  S. Kobayashi, K. Nomizu, Foundations of Differential Geometry I, 		Interscience, 1963.
\item{[20]}  S. Nojiri, S.D. Odintsov,  Conformal anomaly from $dS/CFT$ 		correspondence, hep-th/0106191.
\item{[21]}  S. Gao, R. M. Wald, Theorems on gravitational time delay and related 		issues, Class.Quant.Grav. 17(2000) 4999-5008, gr-qc/0007021.
\item{[22]}  T. Banks, Cosmological Breaking of Supersymmetry?, hep-th/0007146.
\item{[23]}  M. Cai, G. J. Galloway, Boundaries of Zero Scalar Curvature in the 		$AdS/CFT$ Correspondence,
        Adv.Theor.Math.Phys. 3 (1999) 1769, hep-th/0003046.
\item{[24]}  E.Witten, $http://online.itp.ucsb.edu/online/susy\_c99/witten/$
\item{[25]}  G. Galloway, K. Schleich, D. Witt, E. Woolgar, The $AdS/CFT$ 			Correspondence Conjecture and Topological Censorship, Phys.Lett. 			B505 (2001) 255, hep-th/9912119.
\item{[26]}  A. Ashtekar, A. Magnon, Asymptotically Anti-deSitter Space-Times, 			Class. Quant. Grav. 1 (1984) L39.
\item{[27]} N. Seiberg, E. Witten, The D1/D5 System and Singular CFT, JHEP 9904 			(1999) 017, 
     hep-th/9903224.
\item{[28]} C. Baccigalupi, A. Balbi, S. Matarrese, F. Perrotta, N. Vittorio, Constraints on flat cosmologies with tracking Quintessence from Cosmic Microwave Background observations, astro-ph/0109097.
\item{[29]} J. Levin, Topology and the Cosmic Microwave Background, gr-qc/0108043.
\item{[30]}  M. Natsuume, The singularity problem in string theory, gr-qc/0108059.
\item{[31]}  R. M. Schoen,  S-T Yau, On the proof of the postive mass conjecture 			in general relativity,
         Commun. Math.Phys. 65 (1979) 45.
\item{[32]}  R. M. Wald, General Relativity, University of Chicago Press, 1984.
\item{[33]}  M. Spradlin, A. Strominger, A. Volovich, Les Houches Lectures on De Sitter Space, hep-th/0110007.
\item{[34]}  S. Cacciatori, D. Klemm, The Asymptotic Dynamics of de Sitter Gravity in three Dimensions, hep-th/0110031.
\item{[35]}  T. Shiromizu, D. Ida, T. Torii, Gravitational energy, dS/CFT correspondence and cosmic no-hair, hep-th/0109057.
\item{[36]}  C.M. Hull,  De Sitter Space in Supergravity and M Theory, hep-th/0109213.
\item{[37]}  P.O. Mazur, E. Mottola, Weyl Cohomology and the Effective Action for Conformal Anomalies, hep-th/0106151.
\item{[38]}  D. Klemm, Some Aspects of the de Sitter/CFT Correspondence, hep-th/0106247.
\item{[39]}  S. Nojiri, S.D. Odintsov, S. Ogushi, Cosmological and black hole brane-world Universes in higher derivative gravity, hep-th/0108172.
\bye